\documentclass{aa501}
\usepackage{amsmath,graphics,amssymb}

\allowdisplaybreaks

\newfont{\eufont}{eufm10}
\def\eu #1{\mbox{\eufont #1}}

\newcommand{\vr}{{v_r}}
\newcommand{\pr}[1]{{#1}_{\mathrm{p}}}
\newcommand{\io}[1]{{#1}_{\mathrm{i}}}
\newcommand{\zav}[1]{\left(#1\right)}
\newcommand{\hzav}[1]{\left[#1\right]}
\newcommand{\xa}{\io{\eu Y}}
\newcommand{\vth}{v_{\mathrm{th}}}
\newcommand{\prio}[1]{{#1}_{\rm pi}}

\newcommand{\pd}[2]{\frac{\partial #1}{\partial #2}}
\newcommand{\rad}{\mathrm{rad}}
\newcommand{\epr}{{\rm e}^{i\zav{\pr\omega t-\pr\kappa r_{\mathrm{i}}}}}
\newcommand{\eio}{{\rm e}^{i\zav{\io\omega t-\io\kappa r_{\mathrm{i}}}}}
\newcommand{\zm}[1]{#1}
\newcommand{\zmd}[1]{{\bf #1}}
\newcommand{\partialgrad}{\partial_{v'} g^{\rad}}

\begin{document}

\title{Multicomponent radiatively driven stellar winds}
\subtitle{III. Radiative-acoustic waves in a two-component wind}
\titlerunning{Multicomponent radiatively driven stellar winds III.}
%Ku12: --- {\sl (draft version \today)}}

\author{Ji\v{r}\'{\i}  Krti\v{c}ka\inst{1,2},
        Ji\v{r}\'{\i} Kub\'at\inst{2}}
\authorrunning{J. Krti\v{c}ka, J. Kub\'at}

\offprints{J. Krti\v{c}ka,\\ \email{krticka@physics.muni.cz}}

\institute{\'Ustav teoretick\'e fyziky a astrofyziky P\v{r}F MU,
            Kotl\'a\v{r}sk\'a 2, CZ-611 37 Brno, Czech Republic 
           \and
           Astronomick\'y \'ustav, Akademie v\v{e}d \v{C}esk\'e
           republiky, CZ-251 65 Ond\v{r}ejov, Czech Republic}

%\date{draft version \today}
\date{Received 5 December 2001}

\abstract{
We study stability of isothermal two-component
radiatively driven stellar winds against one-dimensional perturbations
larger than the Sobolev length, 
and radiative-acoustic waves in such stellar winds.
\zm{We perform linear perturbation analysis
in comoving fluid-frames of individual
components and 
%Kr13: apply Galilean transformation to 
obtain dispersion
relation in the common fluid-frame.}
For high density winds the velocity difference between velocities of
both components is relatively small and the wind is stable for
radiative-acoustic waves discovered
\zm{originally}
by Abbott, in accordance with the
previous studies of the one-component wind.
However, for such high density winds we found new types of waves
including a special case of 
\zm{"frozen-in" wavy patterns.}
On the other hand, if the velocity difference between wind components is
sufficiently large (for low density winds) then the multicomponent
stellar wind is unstable even for large scale perturbations
\zm{and ion runaway occurs.}
Thus, isothermal two-component stationary solutions of the radiatively
line driven stellar wind with an abrupt lowering of the velocity
gradient are unstable.
   \keywords{stars:   mass-loss  --  
             stars:  early-type  --  
             hydrodynamics
             -- instabilities -- waves}
}

\maketitle
%________________________________________________________________

\section{Introduction}

Since the foundation of the basic theory of the radiatively driven
stellar wind the wind stability was one of the most fundamental
questions to be solved.
At the very beginning Lucy \& Solomon (\cite{LS70}) concluded that
radiatively driven stellar winds are essentially unstable.
Contrary to this assertion,
Castor, Abbott \& Klein (\cite{cak}) described radiatively driven
stellar wind as smooth and stable steady-state outflow.
This contradiction survived when Abbott (\cite{abb}) showed that the
stellar wind described by CAK is stable.
On the other hand, MacGregor et al. (\cite{nesmac}) and
Carlberg (\cite{nesta})
concluded that radiatively driven stellar winds are unstable.
This paradox was solved by Owocki \& Rybicki (\cite{ornest}).
These authors found a general relation which is valid 
for perturbation both smaller and larger than the Sobolev length, the so
called ``bridging relation''.
The main result of the latter paper is that the flow of the line driven
wind is stable for perturbations larger than the Sobolev length 
(the so-called large scale perturbations),
yielding stable radiative-acoustic waves, which were found by Abbott
(\cite{abb}) and unstable for perturbations smaller than the Sobolev
length, as found by MacGregor et al. (\cite{nesmac}) and Carlberg
(\cite{nesta}).
The theory of instabilities of radiatively driven stellar wind was
further developed by Lucy (\cite{L84}), Owocki \& Rybicki (\cite{or2},
\cite{or3}, \cite{or5}), and extended to three-dimensional perturbations
by Rybicki et al. (\cite{or4}).
For an introduction to the problem of a stability of a line driven wind
see Rybicki ({\cite{rybinst}) and Owocki ({\cite{aetso}).
The existence of instabilities in the wind is important for an X-ray
phenomenon because it is usually assumed that X-rays are generated by
wind
clumping or
shocks (e.g. Lucy \& White \cite{luw}, Lucy \cite{L82}, 
Owocki \& Cohen \cite{owco}).

On the other hand, it is
known that radiatively driven stellar
winds have multicomponent nature (e.g., Springmann \& Pauldrach
\cite{treni}, Babel \cite{babela}, Porter \& Drew \cite{iontdisk}). 
The stellar radiation is predominantly absorbed by species (typically C,
N, O, Fe, etc.) which have much lower density than the rest of the
stellar wind, which is composed mainly of hydrogen and helium.
However, multicomponent effects are important only for low-density
stellar winds.
Recently, Krti\v{c}ka \& Kub\'{a}t (\cite{kk}, \cite{kki}, \cite{kkii},
hereafter KK0, KKI, KKII, respectively) computed models of isothermal
two-component and non-isothermal three-component radiatively driven
stellar winds. 

Springmann \& Pauldrach (\cite{treni}) proposed that for low density
stellar winds the absorbing component is not able to accelerate the
non-absorbing component and that both components decouple.
On the other hand, using a model of an isothermal two-component stellar
wind KK0 obtained a surprising result that the components do not
decouple and an unexpected decrease of the velocity gradient was found.
This effect can be explained by the dependence of the radiative force on
the velocity gradient.
A question which naturally arises is the stability of such
multicomponent flow.

It is natural to expect that the conclusions about the stability of the
one-component wind will be in principal also valid for the two component
flows.
Recently, Owocki \& Puls (\cite{op}, hereafter OP) extended the
general one-component stability analysis of Owocki \& Rybicki
(\cite{ornest}) for the case of a two-component isothermal wind and
found that the two-component solution is unstable when the flow is not
well coupled.
Here
we extend Abbott's (\cite{abb})
calculations to the case of the multicomponent flow and study how the
overall picture of stable Abbott waves changes in the two-component
isothermal stellar wind.
The analysis presented in this paper is based on a part of a thesis of
Krti\v{c}ka (\cite{disertacka}).

\section{Time-dependent hydrodynamic equations for isothermal wind}

We assume an isothermal spherically symmetric wind consisting of two
components, namely of passive (non-absorbing) hydrogen ions with mass
equal to proton mass $\pr{m}$ and charge equal to proton charge $\pr{q}$
and of absorbing ions with mass $\io{A}\pr{m}$ and charge $\io{q}$.
Time-dependent radiatively driven stellar wind is then described by the
set of hydrodynamic equations, namely with the continuity equations
\begin{subequations}
\begin{eqnarray}
\label{kontdvoupr}
\pd{\pr{\rho}}{t}+\frac{1}{r^2}\pd{}{r}\zav{r^2\pr{\rho}\pr{\vr}} & = & 0, \\
\pd{\io{\rho}}{t}+\frac{1}{r^2}\pd{}{r}\zav{r^2\io{\rho}\io{\vr}} & = & 0, 
\label{kontdvouio}
\end{eqnarray}
\end{subequations}
and with the equations of motion 
\begin{subequations}
\begin{eqnarray}
\label{pohrovdvoupr}
\pd{\pr{\vr}}{t}+\pr{\vr}\frac{\partial \pr{\vr}}{\partial r}&=&
      -g+\frac{1}{\pr{\rho}}\prio{R}
      -\frac{1}{\pr{\rho}}\frac{\partial {\pr{p}}}{\partial r}, \\
\pd{\io{\vr}}{t}+\io{\vr}\frac{\partial \io{\vr}}{\partial r}&=&
\io{g}^{\mathrm{rad}}-g-\frac{1}{\io{\rho}}\prio{R}
            -\frac{1}{\io{\rho}}\frac{\partial {\io{p}}}{\partial r}.
\label{pohrovdvouio}
\end{eqnarray}
\end{subequations}
In these equations $\pr{\vr}$, $\pr{\rho}$, $\io{\vr}$, $\io{\rho}$ are 
velocities and densities of passive plasma and accelerated ions,
respectively,
$\pr{p}$, $\io{p}$ are partial gas pressures of each component
($\pr{p}=\pr{a}^2\pr{\rho}$,
$\io p=\io{a}^2\io{\rho}$),
isothermal sound velocities are $\pr{a}^2={kT/\pr{m}}$ and
$\io{a}^2={kT/\io{m}}$,
$g=G{\eu M}(1-\Gamma)/r^2$ is the gravitational acceleration (corrected
for absorption
by
free electrons) acting on each component ($G$ and
{\eu M} are gravitational constant and stellar mass, respectively,
$\Gamma$ is the Eddington factor accounting for the absorption on free
electrons) and $\io{g}^{\mathrm{rad}}$ is the radiative acceleration
acting on absorbing ions.
We take the radiative acceleration in the form
\begin{equation}\label{radaccel}
\io{g}^{\mathrm{rad}}= \frac{1}{\xa} \frac{\sigma_e L}{4\pi r^2 c} f
  \zav{\frac{n_e /W }{10^{11}\mbox{cm}^{-3}}}^{\delta}
  \! k\zav{\frac{\xa}{\sigma_e\vth\io{\rho}} \frac{d\io{\vr}}{dr} }^{\alpha}\!,
\end{equation}
with force multipliers $k$, $\alpha$, $\delta$ after Abbott
(\cite{abpar}).
Here $f$ is the finite disk correction factor
(Pauldrach et al. \cite{ppk}, Friend \& Abbott \cite{fa}),
$n_e$ is the electron density (we
set
$n_e=\pr{n}$), and $W$ is stellar dilution factor.
Here we have introduced the factor $\xa$ (which is the ratio of the
absorbing ions density to the passive plasma density in a stellar
atmosphere) to account for radiative acceleration acting directly on
ions (see KK0).

Frictional force (per unit volume) $\prio{R}$ acting between both
components has following form (Springmann \& Pauldrach \cite{treni}):
\begin{equation}\label{Rpi}
\prio{R}=-\pr{n}\io{n}\prio{k}G(\prio{x}),
\end{equation}
where $\pr{n}$ and $\io{n}$ are number densities of passive plasma and
absorbing ions.
The friction coefficient $\prio{k}$ is given by
\begin{equation}
\label{trkcgs}
\prio{k}=\frac{4\pi \ln\Lambda \pr{q}^2\io{q}^2}{k T}
         \frac{\pr{\vr}-\io{\vr}}{|\pr{\vr}-\io{\vr}|},
\end{equation}
where $\ln\Lambda$ is the Coulomb logarithm,
$G(x)$ is the so-called Chandrasekhar function
(see Springmann \& Pauldrach \cite{treni}, KK0 and Fig.~\ref{b5chandra}
for the run of this function).
The argument $\prio{x}$ of the Chandrasekhar function in Eq.~\eqref{Rpi}
is proportional to the ratio of the drift velocity
$\left|\pr{\vr}-\io{\vr}\right|$ to the thermal velocity $\vth$, namely
\begin{equation}
\label{defxpi}
\prio{x}=\sqrt{\prio{A}}\;\frac{\left|\pr{\vr}-\io{\vr}\right|}{\vth},
\end{equation}
where $\prio{A}=\pr{A}\io{A}/\zav{\pr{A}+\io{A}}$
is a reduced atomic mass.

\section{Radiative-acoustic waves}

Similarly to Abbott (\cite{abb}) we keep the equations locally linear
and we study waves in comoving fluid frames of individual components.
Since the velocities of the components are different, we
start with using
two
different fluid frames (one for each component).
Thus, we use comoving fluid frames of non-absorbing and absorbing
components
\begin{subequations}
\label{defpitrans}
\begin{align}
\pr r&=r'-\pr v(r')t, \\
\io r&=r'-\io v(r')t, 
\end{align}
\end{subequations}
instead of the frame of the static observer $r'$ (note that both
comoving fluid-frames are chosen to be local inertial fluid frames).
Here both $\pr r$ and $\io r$ are radial coordinates.
We assume that the wind is perturbed from its original steady-state and
that the perturbed quantities do not change the density scale height.
The perturbed quantities are denoted by
%Ku12:
%$\delta\rho_{a}$, $\delta v_{a}$, where $a$ stands for $\mathrm{p}$
%and $\mathrm{i}$.
$\delta\pr\rho$, $\delta\pr v$, $\delta\io\rho$, and $\delta\io v$.

The time-dependent continuity equations to the first order are
\begin{subequations}
\begin{align}
\label{vlnkontp}
\frac{\partial\delta\rho_{\mathrm{p}}}{\partial t_{\mathrm{p}}}+
\rho_{0,\mathrm{p}}\frac{\partial\delta v_{\mathrm{p}}}{\partial
r_{\mathrm{p}}}&=0,\\
\label{vlnkonti}
\frac{\partial\delta\rho_{\mathrm{i}}}{\partial t_{\mathrm{i}}}+
\rho_{0,\mathrm{i}}\frac{\partial\delta v_{\mathrm{i}}}{\partial
r_{\mathrm{i}}}&=0,
\end{align}
\end{subequations}
and two-component time-dependent linearized momentum equations are
\begin{subequations}
\begin{align}
\frac{\partial\delta v_{\mathrm{p}}}{\partial t_{\mathrm{p}}}&=
-\frac{\pr{a}^2}{\rho_{0,\mathrm{p}}}\frac{\partial\delta\rho_{\mathrm{p}}}{\partial
r_{\mathrm{p}}}+
\label{vlnmomp}
\frac{\prio{R}}{\rho_{0,\mathrm{p}}}\frac{G'(\Delta v_0)}{G(\Delta v_0)}
\zav{\delta v_{\mathrm{i}}-\delta v_{\mathrm{p}}},\\
\frac{\partial\delta v_{\mathrm{i}}}{\partial t_{\mathrm{i}}}&=
-\frac{\io{a}^2}{\rho_{0,\mathrm{i}}}\frac{\partial\delta\rho_{\mathrm{i}}}{\partial
r_{\mathrm{i}}}+
%Kr13: \partial g^{\rad}\frac{\partial\delta
\partialgrad \frac{\partial\delta
v_{\mathrm{i}}}{\partial r_{\mathrm{i}}}-
\nonumber\\*   &-
\frac{\prio{R}}{\rho_{0,\mathrm{i}}}\frac{G'(\Delta v_0)}{G(\Delta v_0)}
\zav{\delta v_{\mathrm{i}}-\delta v_{\mathrm{p}}},
\label{vlnmomi}
\end{align}
\end{subequations}
where the subscript $0$ denotes unperturbed quantity in the observer's
frame, the velocity difference is
$\Delta v_0=v_{0,\mathrm{i}}-v_{0,\mathrm{p}}$, 
$G(\Delta v_0)=G(x_\mathrm{ip})$,
$G'(\Delta v_0)=\partial G(\Delta v_0)/  \partial \Delta v_0$ and
%Kr13 $\partial g^{\rad}= \partial g_{i}^{\mathrm{rad}}/
$\partialgrad= \partial g_{i}^{\mathrm{rad}}/
  \partial \zav{ \partial v_{0,\mathrm{i}} / \partial r_{\mathrm{i}}}$.
We neglected gravity and density stratification in the momentum equation
and assumed that the radiative force depends only on velocity gradient.
In these equations we simply suppose Galilean transformation of
coordinates for which $\pr t = \io t = t$.
We distinguish between $\pr t$ and $\io t$ to emphasize difference in
partial derivatives with respect to time.
During calculation of $\partial/\partial \pr t$ one should keep $\pr r$
constant and, similarly, for $\partial/\partial \io t$
one should keep $\io r$ constant.
The Galilean transformation also implies that the velocity difference
$\Delta v_0$ and perturbations of velocities and densities
$\delta\pr\rho$, $\delta\io\rho$, $\delta \pr v$ a $\delta\io v$ are the
same in both inertial frames.
We neglected density perturbations of the radiative and frictional
forces.
This is consistent with the calculations of Abbott~(\cite{abb}) where
density perturbations of the radiative force have been also neglected.
Justification of this neglect can be found, e.g., in the Appendix of OP.

%Ku11: trosku jsem to preformuloval
%Taking a partial derivative of the equations (\ref{vlnkontp},
%\ref{vlnkonti}) with respect to
%\zm{$\pr r$ and $\io r$ respectively}
%and substituting it to the
%partial derivative of equations (\ref{vlnmomp}, \ref{vlnmomi}) with
%respect to
%\zm{$\pr t$ and $\io t$ respectively}
%gives a system of wave equations
%\begin{subequations}
%\begin{align}
%\label{vlnp}
%\frac{\partial^2\delta v_{\mathrm{p}}}{\partial t_{\mathrm{p}}^2}&=
%\pr{a}^2\frac{\partial^2\delta v_{\mathrm{p}}}{\partial r_{\mathrm{p}}^2}+
%\frac{\prio{R}}{\rho_{0,\mathrm{p}}}\frac{G'(\Delta v_0)}{G(\Delta v_0)}
%  \hzav{\frac{\partial\delta v_{\mathrm{i}}}{\partial t_{\mathrm{p}}}-
%       \frac{\partial\delta v_{\mathrm{p}}}{\partial t_{\mathrm{p}}}},\\
%\frac{\partial^2\delta v_{\mathrm{i}}}{\partial t_{\mathrm{i}}^2}&=
%\io{a}^2\frac{\partial^2\delta v_{\mathrm{i}}}{\partial r_{\mathrm{i}}^2}+
%\partial g^{\mathrm{rad}}\frac{\partial^2\delta v_{\mathrm{i}}}{\partial
%r_{\mathrm{i}}\partial t_{\mathrm{i}}}-
%\nonumber\\*  & -
%\frac{\prio{R}}{\rho_{0,\mathrm{i}}}\frac{G'(\Delta v_0)}{G(\Delta v_0)}
%  \hzav{\frac{\partial\delta v_{\mathrm{i}}}{\partial t_{\mathrm{i}}}-
%       \frac{\partial\delta v_{\mathrm{p}}}{\partial t_{\mathrm{i}}}}.
%\label{vlni}
%\end{align}
%\end{subequations}
%
\zm{
Taking a partial derivative of the equation \eqref{vlnkontp} with
respect to $\pr r$ and substituting it to the partial derivative of the
equation \eqref{vlnmomp} with respect to $\pr t$ gives
\begin{subequations}
\begin{align}
\label{vlnp}
\frac{\partial^2\delta v_{\mathrm{p}}}{\partial t_{\mathrm{p}}^2}&=
\pr{a}^2\frac{\partial^2\delta v_{\mathrm{p}}}{\partial r_{\mathrm{p}}^2}+
\frac{\prio{R}}{\rho_{0,\mathrm{p}}}\frac{G'(\Delta v_0)}{G(\Delta v_0)}
  \hzav{\frac{\partial\delta v_{\mathrm{i}}}{\partial t_{\mathrm{p}}}-
       \frac{\partial\delta v_{\mathrm{p}}}{\partial t_{\mathrm{p}}}}.
\intertext{Similarly, partial derivative of the equation
\eqref{vlnkonti} with respect to $\io r$ and subsequent substitution
to the derivated equation \eqref{vlnmomi} with respect to $\io t$ gives}
\frac{\partial^2\delta v_{\mathrm{i}}}{\partial t_{\mathrm{i}}^2}&=
\io{a}^2\frac{\partial^2\delta v_{\mathrm{i}}}{\partial r_{\mathrm{i}}^2}+
%Kr13 \partial g^{\mathrm{rad}}\frac{\partial^2\delta v_{\mathrm{i}}}{\partial
\partialgrad\frac{\partial^2\delta v_{\mathrm{i}}}{\partial
r_{\mathrm{i}}\partial t_{\mathrm{i}}}-
\nonumber\\*  & -
\frac{\prio{R}}{\rho_{0,\mathrm{i}}}\frac{G'(\Delta v_0)}{G(\Delta v_0)}
  \hzav{\frac{\partial\delta v_{\mathrm{i}}}{\partial t_{\mathrm{i}}}-
       \frac{\partial\delta v_{\mathrm{p}}}{\partial t_{\mathrm{i}}}}.
\label{vlni}
\end{align}
\end{subequations}
}

\zm{We obtained two differential equations for velocity perturbations
of both components.
Unfortunately, these equations are written in {\em different} fluid
frames.
To proceed further, we have to rewrite these equations in one common
frame.}
We selected the fluid frame of accelerated ions.
From the relations
\begin{subequations}
\begin{align}
\io r= &\pr r -\Delta v_0 \pr t,\\
\io t= &\pr t,
\end{align}
\end{subequations}
it follows that
\begin{subequations}
\begin{align}
\frac{\partial}{\partial r_{\mathrm{p}}}&=
\frac{\partial}{\partial r_{\mathrm{i}}},\\*
\frac{\partial}{\partial t_{\mathrm{p}}}&=
\frac{\partial}{\partial t_{\mathrm{i}}}-
\Delta v_0 \frac{\partial}{\partial r_{\mathrm{i}}},\\
\frac{\partial^2}{\partial r_{\mathrm{p}}^2}&=
\frac{\partial^2}{\partial r_{\mathrm{i}}^2},\\*
\frac{\partial^2}{\partial t_{\mathrm{p}}^2}&=
\frac{\partial^2}{\partial t_{\mathrm{i}}^2}-
2\Delta v_0 \frac{\partial^2}{\partial t_{\mathrm{i}} \partial r_{\mathrm{i}}}+
\Delta v_0^2 \frac{\partial^2}{\partial r_{\mathrm{i}}^2}.
\end{align}
\end{subequations}
Thus, we rewrite the Eq.~(\ref{vlnp}) as
\begin{multline}
\frac{\partial^2\delta v_{\mathrm{p}}}{\partial t_{\mathrm{i}}^2}
-2\Delta v_0 \frac{\partial^2\delta v_{\mathrm{p}}}{\partial t_{\mathrm{i}}
\partial r_{\mathrm{i}}}+
\Delta v_0^2\frac{\partial^2\delta v_{\mathrm{p}}}{\partial r_{\mathrm{i}}^2}=
\pr{a}^2\frac{\partial^2\delta v_{\mathrm{p}}}{\partial r_{\mathrm{i}}^2}+\\*+
\frac{\prio{R}}{\rho_{0,\mathrm{p}}}\frac{G'(\Delta v_0)}{G(\Delta v_0)}
\zav{\frac{\partial}{\partial t_{\mathrm{i}}}-
           \Delta v_0 \frac{\partial}{\partial r_{\mathrm{i}}}}
  \zav{\delta v_{\mathrm{i}}-
       \delta v_{\mathrm{p}}}.
\label{upvlnpp}
\end{multline}

We assume a solution in the form of propagating waves, which in the
reference frame of ions 
\zm{are}
\begin{subequations}\label{delva}
\begin{align}
\delta \pr v&=\pr V \exp
\hzav{i\zav{\pr\omega t_{\mathrm{i}}-\pr\kappa r_{\mathrm{i}}}},\\
\delta \io v&=\io V \exp
\hzav{i\zav{\io\omega t_{\mathrm{i}}-\io\kappa r_{\mathrm{i}}}}.
\end{align}
\end{subequations}
The 
\zm{amplitudes $\pr V$, $\io V$ are}
generally complex to account for phase shifts,
and, since we are doing linear analysis,
\zm{they do}
depend neither on
$\io r$ nor on $\io t$.
Similarly, 
\zm{$\pr\omega$, $\io\omega$, $\pr\kappa$, and $\io\kappa$}
are independent of $\io r$ and
$\io t$.
Substituting from (\ref{delva}) into the wave equations (\ref{upvlnpp},
\ref{vlni}) we obtain a system of equations 
\begin{subequations}
\label{dispexp}
\begin{multline}
\label{dispexppr}
\left[\zav{\pr\omega+\Delta v_0\pr\kappa}^2\pr V-
\pr{a}^2\pr{\kappa}^2\pr V -\right.  \\* \left. -
i \frac{\prio{R}} {\rho_{0,\mathrm{p}}}\frac{G'(\Delta v_0)}{G(\Delta v_0)}
\zav{\pr\omega+\Delta v_0\pr\kappa}\pr V\right]\epr  = \\* =-
i \frac{\prio{R}} {\rho_{0,\mathrm{p}}}\frac{G'(\Delta v_0)}{G(\Delta v_0)}
\zav{\io\omega+\Delta v_0\io\kappa}\io V\eio,
\end{multline}
\begin{multline}
\label{dispexpio}
\left[\io{\omega}^2\io V - \io{a}^2\io\kappa^2\io V +  \io\kappa
%Kr13 \partial g^{\mathrm{rad}}\io\omega \io V - \right. \\* \left. -
\partialgrad\io\omega \io V - \right. \\* \left. -
i \frac{\prio{R}}{\rho_{0,\mathrm{i}}}
\frac{G'(\Delta v_0)}{G(\Delta v_0)}\io\omega\io V\right]\eio = \\* =
-i \frac{\prio{R}}{\rho_{0,\mathrm{i}}}
\frac{G'(\Delta v_0)}{G(\Delta v_0)}\pr\omega\pr V\epr.
\end{multline}
\end{subequations}
We are looking for a non-trivial solution of Eqs.~(\ref{dispexp}), i.e.
a solution with $\io V\neq 0$, $\pr V \neq 0$ for arbitrary $t$,
$\io r$.
Eq.~(\ref{dispexppr}) can be rewritten as
\begin{displaymath}
\widetilde{A}_{\mathrm{pp}}\pr V\epr=
\widetilde{A}_{\mathrm{pi}}\io V\eio
\end{displaymath}
or
\begin{displaymath}
\pr V=\frac{\widetilde{A}_{\mathrm{pi}}}{\widetilde{A}_{\mathrm{pp}}}
\exp\hzav{i\zav{\io\omega-\pr\omega} t-
          i\zav{\io\kappa-\pr\kappa}r_{\mathrm{i}}} \io V.
\end{displaymath}
Because $\pr V$, $\io V$, $\widetilde{A}_{\mathrm{pp}}$ and
$\widetilde{A}_{\mathrm{pi}}$ do not depend on $t$ and $\io r$, the last
equation can be fulfilled only if
\begin{equation}
\label{okpod1}
\omega_{\mathrm{p}}t-\kappa_{\mathrm{p}}r_{\mathrm{i}}=
\omega_{\mathrm{i}}t-\kappa_{\mathrm{i}} r_{\mathrm{i}}.
\end{equation}
holds for any $t$, $\io r$ (the same conclusion can be obtained from
Eq.~(\ref{dispexpio})).
Thus, wavenumbers and frequencies of both components are the same,
%Ku12: zmenime cislovani, cimz recenzenta uplne zmateme :-)
\begin{subequations}
\begin{eqnarray}
\io \kappa & = & \pr \kappa,\\
\io \omega  & = & \pr \omega .
\end{eqnarray}
%Ku12:
\end{subequations}
\zm{Dividing Eqs.~(\ref{dispexp}) by the exponential factor and denoting
$\kappa\equiv \pr \kappa$ and $\omega\equiv\pr \omega$ we can rewrite
the remaining system of equations as}
\begin{subequations}
\label{dispc}
\begin{multline}
\label{disppr}
\zav{\omega+\Delta v_0\kappa}^2\pr V = 
\pr{a}^2\kappa^2\pr V  -  \\* -
i \frac{\prio{R}} {\rho_{0,\mathrm{p}}}\frac{G'(\Delta v_0)}{G(\Delta v_0)}
\zav{\omega+\Delta v_0\kappa}\zav{\io V-\pr V},
\end{multline}
\begin{multline}
\label{dispio}
\omega^2\io V = \io{a}^2\kappa^2\io V -  \kappa
%Kr13 \partial g^{\mathrm{rad}}\omega \io V + \\*   +
\partialgrad\omega \io V + \\*   +
i \frac{\prio{R}}{\rho_{0,\mathrm{i}}}
\frac{G'(\Delta v_0)}{G(\Delta v_0)}\omega\zav{\io V-\pr V}.
\end{multline}
\end{subequations}

This system of equations can be rewritten in the matrix form as
\begin{equation}
\label{dispmatrov}
\mathsf{A}\mathbf{V}=0,
\end{equation}
where the vector $\mathbf{V}=(\pr V,\io V)^\mathrm{T}$
and individual elements of the matrix $\mathsf{A}$ are
\begin{subequations}
\begin{align}
\label{app}
A_{\mathrm{pp}}&=\omega^2+2\Delta v_0\kappa\omega+\Delta v_0^2\kappa^2- \nonumber
        \pr{a}^2\kappa^2 -\\*& - i\pr P \zav{\omega+\Delta v_0\kappa},\\
\label{api}
A_{\mathrm{pi}}&=i\pr P \zav{\omega+\Delta v_0\kappa},\\
A_{\mathrm{ip}}&=i \io P \omega,\\
%Kr13: A_{\mathrm{ii}}&=\omega^2-\io{a}^2\kappa^2+ \kappa \partial g^{\mathrm{rad}}\omega-
A_{\mathrm{ii}}&=\omega^2-\io{a}^2\kappa^2+ \kappa \partialgrad\omega-
       i \io P \omega,
\end{align}
\end{subequations}
where
\begin{subequations}
\begin{align}
\pr P=&
\frac{\prio{R}}{\rho_{0,\mathrm{p}}} \frac{G'(\Delta v_0)}{G(\Delta v_0)},\\
\io P=&
\frac{\prio{R}}{\rho_{0,\mathrm{i}}} \frac{G'(\Delta v_0)}{G(\Delta v_0)}.
\end{align}
\end{subequations}
Equation (\ref{dispmatrov}) has a non-zero solution only if
\begin{equation}
\label{disobec}
||\mathsf{A}||=0,
\end{equation}
which is the dispersion relation.
Generally, it has a complicated form and we will solve it numerically.
However, in order to better understand the general dispersion relation
we shall first find an analytical solution for some
simpler
specific cases.

\subsection{Abbott waves}
\label{kapabb}

Let us assume that the velocity amplitudes of the components
$\mathrm{p}$ and $\mathrm{i}$ are nearly equal ($\io V\approx \pr V$)
and that the phase velocity of the wave is much larger than the drift
velocity ($\omega/k\gg \Delta v_0$), i.e. the flow is well coupled.
Then the terms containing $\Delta v_0$ in the Eq.~(\ref{disppr}) 
can be neglected.
These conditions mimic the one-component case, for which Abbott
(\cite{abb}) obtained stable radiative-acoustic waves, the so called
Abbott waves.
We may expect that for well coupled high density winds the two-component
waves are similar to the Abbott waves.
Indeed, summing dispersion relations (\ref{dispc}) the imaginary
frictional term vanishes, and we obtain the dispersion equation in the
form of two-component Abbott waves
\begin{multline}
\label{dispA}
\zav{\omega^2 -\pr{a}^2\kappa^2}\!\rho_{0,\mathrm{p}}+
%Kr13: \zav{\omega^2 + \omega \kappa \, \partial g^{\mathrm{rad}}-
\zav{\omega^2 + \omega \kappa \, \partialgrad-
      \io{a}^2\kappa^2 }\rho_{0,\mathrm{i}}=0.
\end{multline}
This equation corresponds to the one-component dispersion relation and
thus justifies the condition used by KK0 to fix the mass-loss rate.
Then solving the Eq.~(\ref{dispA}) for $\omega$, the dispersion relation
takes the form of
\begin{multline}
\label{abbvln}
\omega=\kappa\hzav{
-\frac{1}{2}\frac{\rho_{0,\mathrm{i}}}{\rho_{0,\mathrm{p}}}
%Kr13                     \partial g^{\mathrm{rad}} \pm
                     \partialgrad \pm
   \sqrt{\zav{\frac{1}{2}\frac{\rho_{0,\mathrm{i}}}{\rho_{0,\mathrm{p}}}
%Kr13          \partial g^{\mathrm{rad}}}^2+\pr{a}^2 }},
          \partialgrad}^2+\pr{a}^2 }},
\end{multline}
where we neglected density of absorbing ions compared to the passive
plasma density.
Note that the Eq.~(\ref{abbvln}) is the same as the Eq.~(47) of Abbott
(\cite{abb}), since we used a modified definition of the driving force
Eq.~(\ref{radaccel}) to account for the fact that only the ionic gas
is line driven.
Apparently, $\omega$ is real and thus this mode is neither unstable nor
damped.
There are two branches of $\omega$ corresponding to forward and backward
waves, both depending on a wavenumber linearly.
Note, however, that the general dispersion relation derived by Owocki \&
Rybicki (\cite{ornest}) for the one-component flow allows for
instabilities for short-wavelength perturbations.

The point where the velocity of backward waves is equal to the wind
velocity is called the critical point.
The flow above the critical point cannot communicate with the wind base.
Thus, early type stars are surrounded by the critical surface which
separates two different domains of the stellar wind.
Feldmeier \&  Shlosman (\cite{feslo00})
aptly
compare this situation to the cosmic censorship hypothesis.
The result that the multicomponent nature of the wind does not alter the
Abbott waves is important because these waves determine the mass-loss of
the CAK wind (Feldmeier \& Shlosman \cite{feslo01}).

Finally, in the case when the gas-pressure term $\pr{a}^2$ can be
neglected, the dispersion relation (\ref{abbvln}) corresponds to the
$\omega_+$ mode of OP, namely
\begin{equation}
\omega_+\approx-\kappa \frac{\rho_{0,\mathrm{i}}}{\rho_{0,\mathrm{p}}}
%Kr13: \partial g^{\mathrm{rad}}.
\partialgrad.
\end{equation}
%Kr13: \zm{Here, the $+$ subscript indicates downstream propagation of waves.}
\zmd{We denoted this mode in accordance with OP as $\omega_+$ although
this mode is upstream.
Note that Eq.(\ref{abbvln}) has another solution for
%Ku13: $\pr{a}^2=0$,
$\pr{a}=0$,
namely $\omega = 0$.}

\subsection{Purely ionic Abbott waves}
\label{kapiontvln}

\begin{figure}
\resizebox{\hsize}{!}{\includegraphics{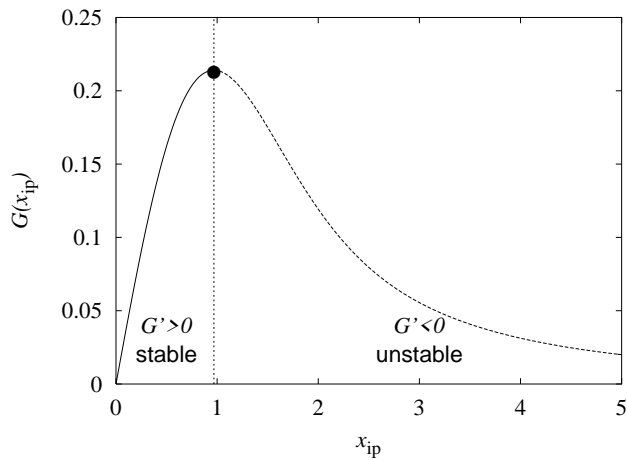}}
\caption[]{The run of Chandrasekhar function
(note, that Eq.~(\ref{defxpi}) yields $x_{\mathrm{ip}}\sim\Delta v_0$).
When the flow is well coupled, $x_{\mathrm{ip}}\lesssim 0.97$,
$G'(\Delta v_0)>0$ and the wind is stable.
When the drift velocity is large, $x_{\mathrm{ip}}\gtrsim 0.97$,
$G'(\Delta v_0)<0$ and the wind is unstable.
Note that the point $x_{\mathrm{ip}} \approx 0.97$ corresponds to the
maximum of $G$.}
\label{b5chandra}
\end{figure}

In the case when $|\io V|\gg |\pr V|$ we can obtain from the
Eq.~(\ref{dispio}) the dispersion relation in the form of Abbott waves
in absorbing ions
\begin{equation}
\label{dispohrom}
\omega^2=\io{a}^2\kappa^2-\omega
\kappa
%Kr13: \, \partial g^{\mathrm{rad}}+
\, \partialgrad+
i \frac{\prio{R}}{\rho_{0,\mathrm{i}}}
\frac{G'(\Delta v_0)}{G(\Delta v_0)}\omega.
\end{equation}

We will study such waves in the case when the pressure term
$\io{a}^2\kappa^2$ in the dispersion equation (\ref{dispohrom}) can be
neglected.
In this case, the dispersion relation has a form,
which corresponds to the $\omega_-$ mode of OP,
\begin{equation}
\label{enabb}
%Kr13: \omega_-=-\kappa \, \partial g^{\mathrm{rad}}+
\omega_-=-\kappa \, \partialgrad+
i \frac{\prio{R}}{\rho_{0,\mathrm{i}}}
\frac{G'(\Delta v_0)}{G(\Delta v_0)},
\end{equation}
%Kr14 \zm{where the $-$ subscript means upstream propagation of waves.}
The real part of
$\omega_-$
depends on $\kappa$ linearly, whereas the
imaginary part does not depend on $\kappa$.
Due to the presence of the imaginary term in Eq.~(\ref{enabb}) such
waves are damped in the case when $G'(\Delta v_0)>0$.
\zm{For clearness we plotted the run of the Chandrasekhar function in
Fig.~\ref{b5chandra}}
Thus, if the wind is coupled
[when
the drift velocity is lower than the
thermal velocity or, more precisely, when the Chandrasekhar function is
a rising function of $\Delta v_0$ ($G'(\Delta v_0)>0$)],
the wind is stable for this type of waves.
On the other hand, when the Chandrasekhar function $G(x)$ is decreasing
($G'(\Delta v_0)<0$), the wind is not stable for the above mentioned
perturbations,
\zm{and this mode leads to an ion runaway.}

Physical reasons for such instability are straightforward.
If the Chandrasekhar function is before its maximum (as function of
$\Delta v_0$), then the increase of the velocity difference
between velocities of ions and passive plasma
enhances the frictional force, which finally tends to lower the velocity
difference yielding a stable flow.
In the opposite case (if the velocity difference is larger than that
corresponding to the maximum of Chandrasekhar function) the increase of
the velocity difference lowers the frictional force which allows for
additional increase of the velocity difference.
Such two-component flow is clearly unstable.
\zm{This is the effect of the ion runaway
(see also Springmann \& Pauldrach \cite{treni}, OP).}

At a first glance there might exist similar acoustic waves in the
non-absorbing component.
However, this is not the case, because the imaginary term in the 
Eq.~(\ref{dispio}) is larger than in Eq.~(\ref{disppr}) and thus
Eq.~(\ref{dispio}) does not allow 
$|\pr V|\gg |\io V|$.
Therefore there are no acoustic counterparts of such waves in a
non-absorbing component, i.e. there are no passive plasma waves for
which the condition $|\pr V|\gg |\io V|$ is valid.

\subsection{Zero frictional force}

\zm{To complete the list of simplified cases, we have also to mention
the hypothetical case when interaction between the components vanishes.}
If both flow components do not influence each other, i.e. if the
frictional force is zero,
then the
system of equations (\ref{dispexp}) does not implicate that the
frequencies and wavenumbers of both components are equal.
Instead of the system of dispersion relations~(\ref{dispc}) we obtain
two independent relations for each wind component
\begin{subequations}
\begin{align}
\label{indepp}
\zav{\pr\omega+\Delta v_0\pr\kappa}^2-
\pr{a}^2\pr{\kappa}^2&=0,\\
\label{indepi}
\io{\omega}^2- \io{a}^2\io\kappa^2 +  \io\kappa
%Kr13: \partial g^{\mathrm{rad}}\io\omega&=0. 
\partialgrad\io\omega&=0. 
\end{align}
\end{subequations}
The dispersion relation of the nonabsorbing component (\ref{indepp})
describes ordinary isothermal sound waves and the dispersion relation of
the ionic component (\ref{indepi}) describes stable Abbott waves of
absorbing ions.
%Kr11:
%Ku11: Tomu nerozumim, jakou mnozinu rovnic mas na mysli? Musi to tu
%      byt?
%\zm{However, in the real plasmas this situation shall be probably
%modelled by more elaborate set of equations.}
%Kr12: Jenom jsem chtel rict to, co zminoval Feldmaier, ze popis
% takoveho plazmatu by asi byl slozitejsi. Vcelku to ale tam byt nemusi.
%

Note that a similar result of independent waves can be obtained for the
maximum of Chandrasekhar function, for which $G'(\Delta v_0)=0$ and the
interaction terms in Eqs.~(\ref{dispc}) vanish as for the case of zero
frictional force.

\section{Numerical results}

The simplified calculations presented in Sections \ref{kapabb} and
\ref{kapiontvln} can help us to better understand the behavior of
individual branches of the general dispersion relation (\ref{disobec}).
These calculations have been done numerically.
To study the individual branches of the dispersion relation we solved
numerically the Eq.~(\ref{disobec}) using the procedure {\tt CPOLY}
of Jenkins \& Traub (\cite{jetra}).

Moreover, solving Eq.~(\ref{disppr}) or Eq.~(\ref{dispio}) for given
$\omega$ and $\kappa$ we obtain the relation between wind amplitudes
\begin{subequations}
\begin{equation}
\label{vivp}
\io V=-\frac{A_{\mathrm{pp}}}{A_{\mathrm{pi}}}\pr V
\end{equation}
or, equivalently,
\begin{equation}
\io V=-\frac{A_{\mathrm{ip}}}{A_{\mathrm{ii}}}\pr V.
\end{equation}
\end{subequations}

\subsection{Stable wind for ionic Abbott waves}
\label{Owaves}

\begin{figure}
\resizebox{\hsize}{!}{\includegraphics{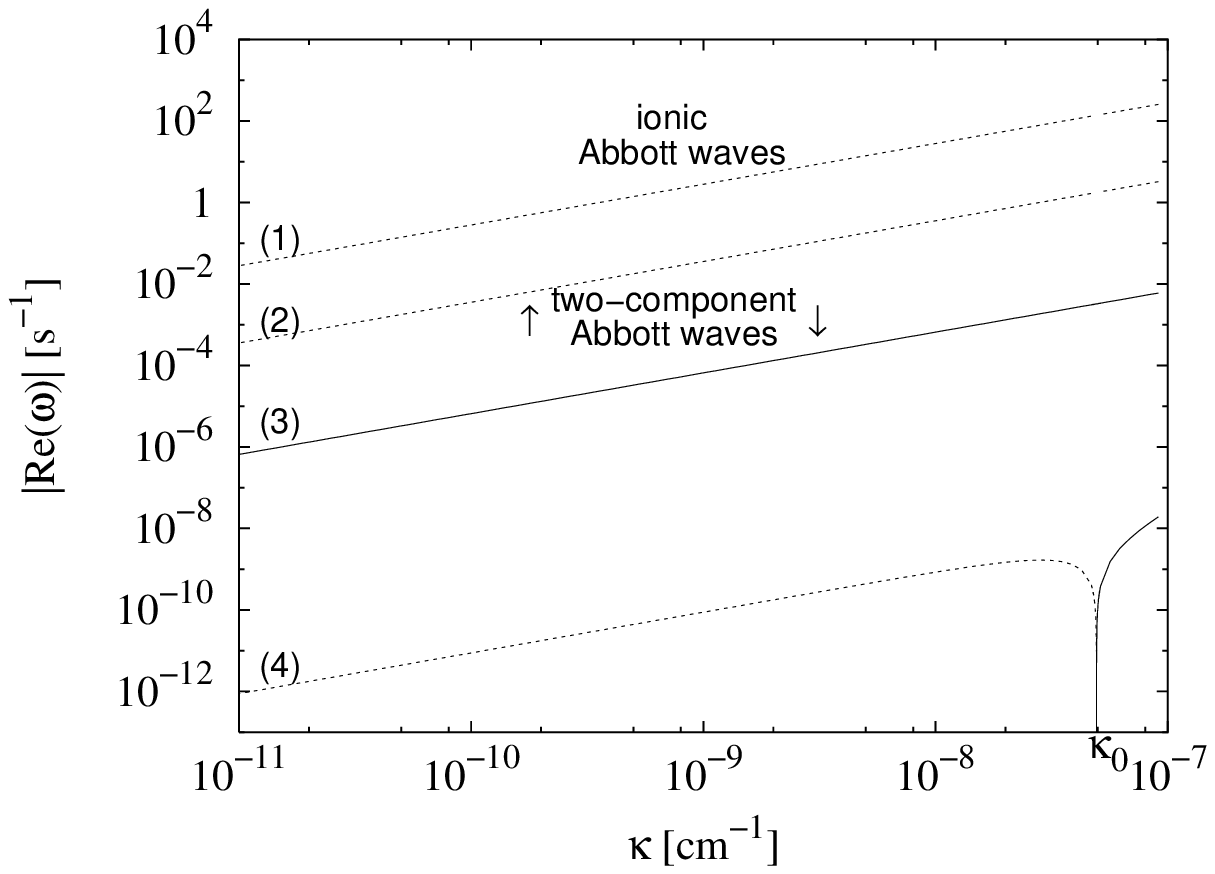}}
\resizebox{\hsize}{!}{\includegraphics{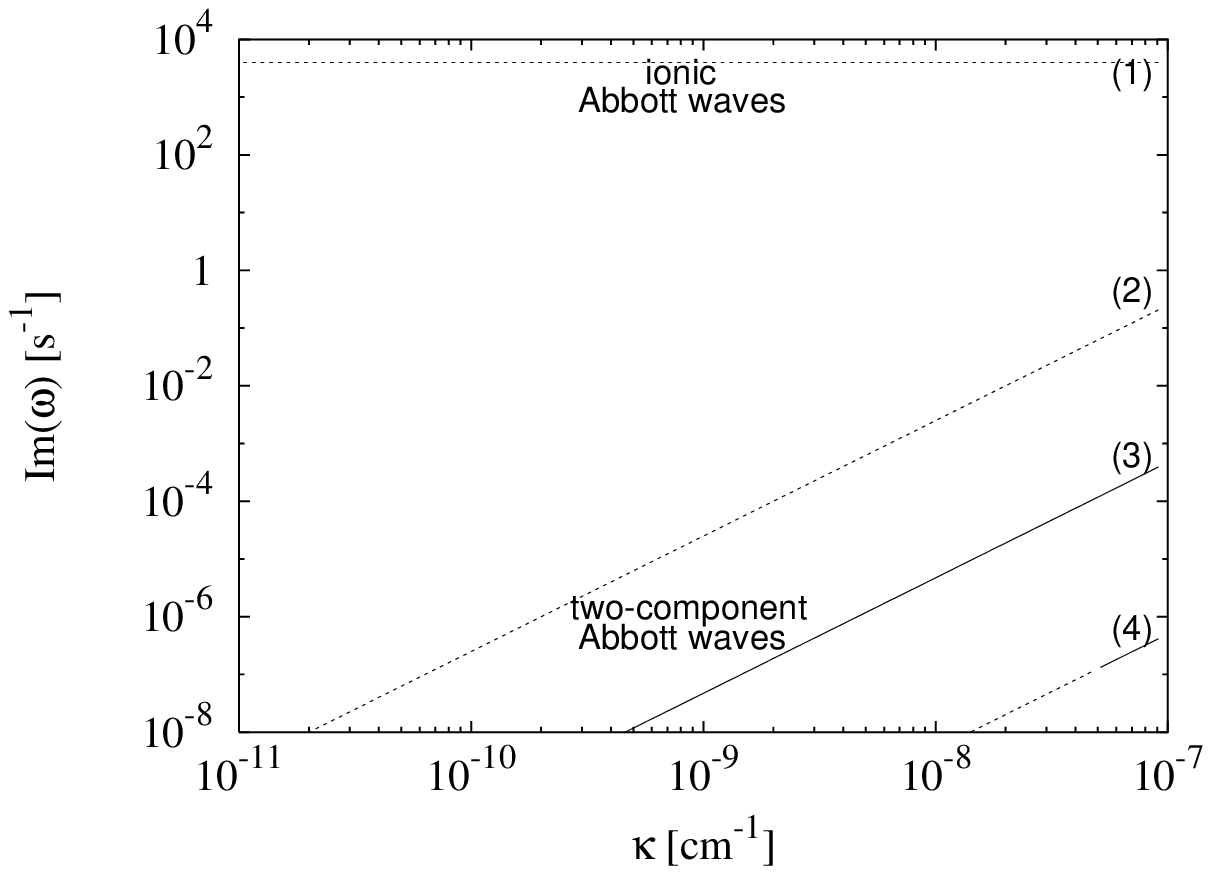}}
\caption[]{Individual branches of the dispersion relation for the point
where $\vr\approx 0.3\,v_{\infty}$ in the
dense wind for a specific case of $\epsilon$~Ori.
{\em Upper panel:} Real roots of $\omega$.
Solid line represents a forward wave, dashed ones backward waves.
Note, that the real part of branch denoted as (4) changes its signature
and for $\kappa=\kappa_0$ is the real part of frequency $\omega$ zero.
{\em Lower panel:} Imaginary roots of $\omega$, all waves are damped.
As in the upper panel,
solid and dashed lines denote forward and backward waves,
respectively.
Real and imaginary parts of individual roots are denoted using the same
numbers.
%Ku13: Ja bych to tam nedaval. Neumim si zduvodnit, proc je to
%      nesmysluplne. To ze je neco male, se da napsat, to se opravdu
%      nemusi malovat.
%Kr13:
%\zmd{Althought quite large range in $\Im(\omega)$ is not neccesseary
%physically meaningfull, we ploted it to show that calculated imaginary
%parts of given roots are very small.}
%
}
\label{dispepsori}
\end{figure}

First, we shall study the case of the star $\epsilon$~Ori described in
KK0.
This star has a relatively dense wind, so the drift velocity between
both components is low compared to the thermal speed of hydrogen and
thus this wind should be stable for ionic Abbott waves because
$G'(\Delta v_0)>0$ in this case (cf. Section \ref{kapiontvln}).
To be specific, we shall study the dispersion relation in the point
where $\vr\approx 0.3\,v_{\infty}$.
This
choice
does not influence the overall picture of our results, however.

Resulting
dispersion relations are displayed in the Fig.~\ref{dispepsori}.
Real parts of branches show linear dispersion relations as
was predicted in
Eqs.~(\ref{abbvln}) and (\ref{enabb}).
The real branches of Abbott waves (two middle straight lines in both
panels of Fig.~\ref{dispepsori}
denoted by (2) and (3))
in the two-component case are
essentially the same as in the one-component case
(cf. Abbott \cite{abb}).
From the corresponding imaginary parts of these two-component Abbott
waves follows that these waves are damped only marginally.
If we extend the calculations to 
$\kappa > 10^{-7}\,\mathrm{cm}^{-1}$
we could in
principle obtain larger damping, but these values are beyond the region
of validity of the assumption of perturbations larger than the Sobolev
length. 
We also calculated the relation between wave amplitudes $\io V$ and
$\pr V$ of absorbing and nonabsorbing components using Eq.~(\ref{vivp}).
This calculation confirmed the assumption used in Sect. \ref{kapabb}
for calculation of two-component Abbott waves that wave amplitudes for
this waves are nearly the same.

Largest imaginary branch corresponding to the ionic Abbott waves (upper
straight line in both panels of Fig.~\ref{dispepsori}) does not depend
on~$\kappa$ (see Eq.~(\ref{enabb})).
Clearly, according to previous results, such waves are heavily damped.
Moreover, the calculation of the relation between $\io V$ and $\pr V$
confirmed that $|\io V|\gg |\pr V|$  for these waves
(cf. Sect. \ref{kapiontvln}).

However, a new type of slow waves appeared.
They are described by the lower curves in both panels of
Fig.~\ref{dispepsori} (denoted as (4)).
The real part of the dispersion relations shows almost linear
dependence with the exception of the region where it passes through the
value of $\Re(\omega)=0$.
Consequently, this solution corresponds to both forward and backward
waves, which are stable and only marginally damped.
The case of $\Re(\omega)=0$ deserves special attention.
It corresponds to a static wavy structure in the comoving frame, so in
the observer frame this structure resembles almost stable outflowing
\zm{"frozen-in" wavy patterns}
of the characteristic size of 
$\kappa_0^{-1}\approx10^7\,\mathrm{cm}$ for this specific case
(note that $\kappa_0$ is wavenumber for which $\Re(\omega)=0$).

In addition,
calculations showed that the value of wavenumber $\kappa_0$ (for which
$\Re(\omega)=0$) depends on the distance from the star.
At the base of the wind the value of $\kappa_0$ is lower,
$\kappa_0\approx 10^{-6}\,\mathrm{cm}^{-1}$ implying the possible
characteristic 
\zm{pattern}
size of the order $10^{6}\,\mathrm{cm}$ whereas in
the outer parts of the wind
$\kappa_0\approx 10^{-10}\,\mathrm{cm}^{-1}$ yielding the
characteristic
\zm{pattern}
size of the order $10^{10}\,\mathrm{cm}$.

For this new type of waves we can obtain approximate analytical
dispersion relation.
Calculations showed that due to the low value of $\omega$ for these
waves all terms in the dispersion relation (\ref{disobec}) can be
neglected except constant terms and terms linear in $\omega$.
Further neglect of all terms which do not significantly influence this
type of waves leads to the dispersion relation in the form of
\begin{equation}
%Kr13: \omega\zav{-\pr{a}^2 \partial g^{\mathrm{rad}}\kappa +i\pr{a}^2\io P}
\omega\zav{-\pr{a}^2 \partialgrad\kappa +i\pr{a}^2\io P}
+\kappa^2\pr{a}^2\io{a}^2+i\Delta v_0\pr P\io{a}^2\kappa \approx 0.
\end{equation}
The condition $\Re(\omega)=0$ can be now written as
\begin{equation}
%Kr13: \kappa^2\pr{a}^2 \partial g^{\mathrm{rad}}-
\kappa^2\pr{a}^2 \partialgrad-
\Delta v_0\pr P \io P \approx 0
\end{equation}
from which we can obtain the equation for $\kappa_0$ in the form of
\begin{equation}
%Kr13: \kappa_0\approx\frac{1}{\pr{a}}\sqrt{\frac{\Delta v_0\pr P \io P }{\partial g^{\mathrm{rad}}}}.
\kappa_0\approx\frac{1}{\pr{a}}\sqrt{\frac{\Delta v_0\pr P \io P }{\partialgrad}}.
\end{equation}

However, we must keep in mind that our analysis was only linear, taking 
into account the nonlinear effects may change this promising picture,
similarly to the case of one-component Abbott waves which become
unstable if the second order effects are taken into account (Feldmeier
\cite{feld98}).

\subsection{Unstable wind for ionic Abbott waves}
\label{kapnesta}

\begin{figure}
\resizebox{\hsize}{!}{\includegraphics{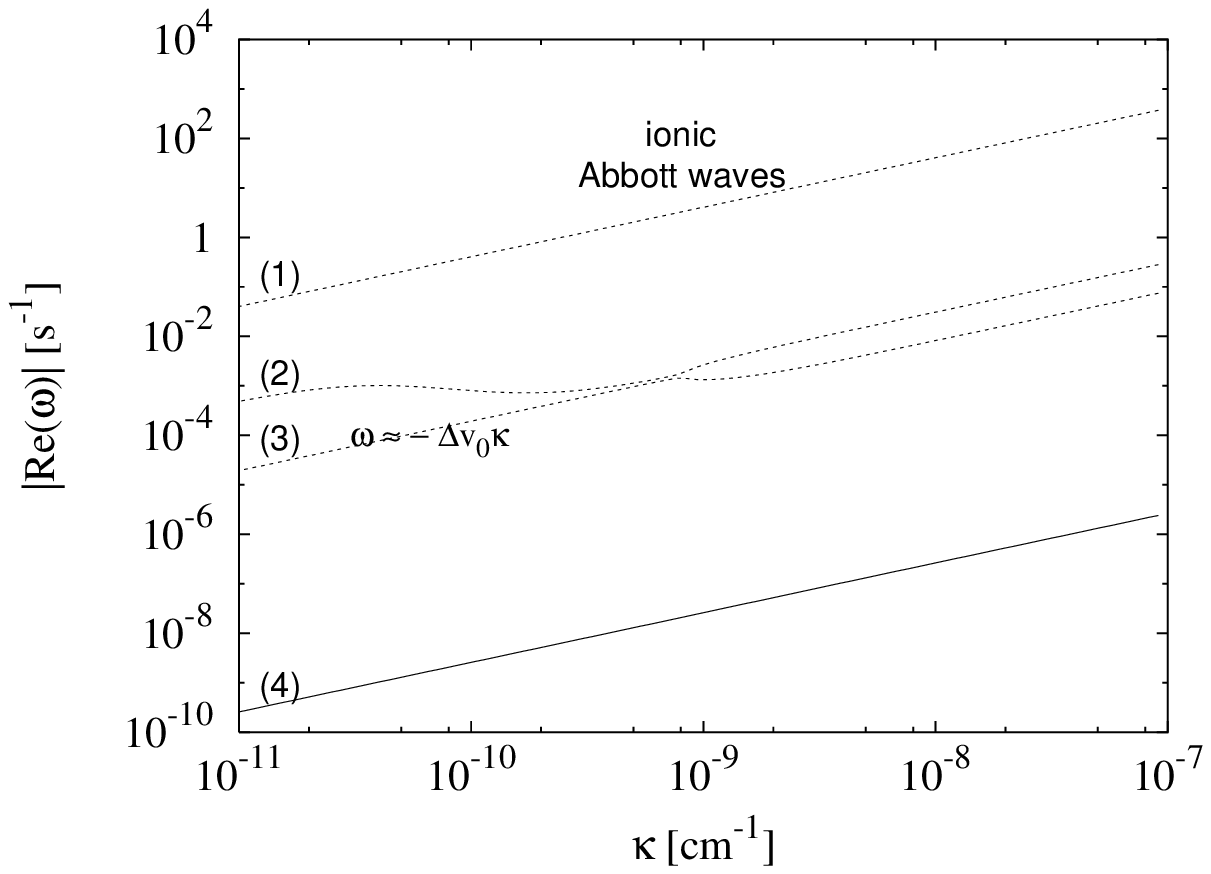}}
\resizebox{\hsize}{!}{\includegraphics{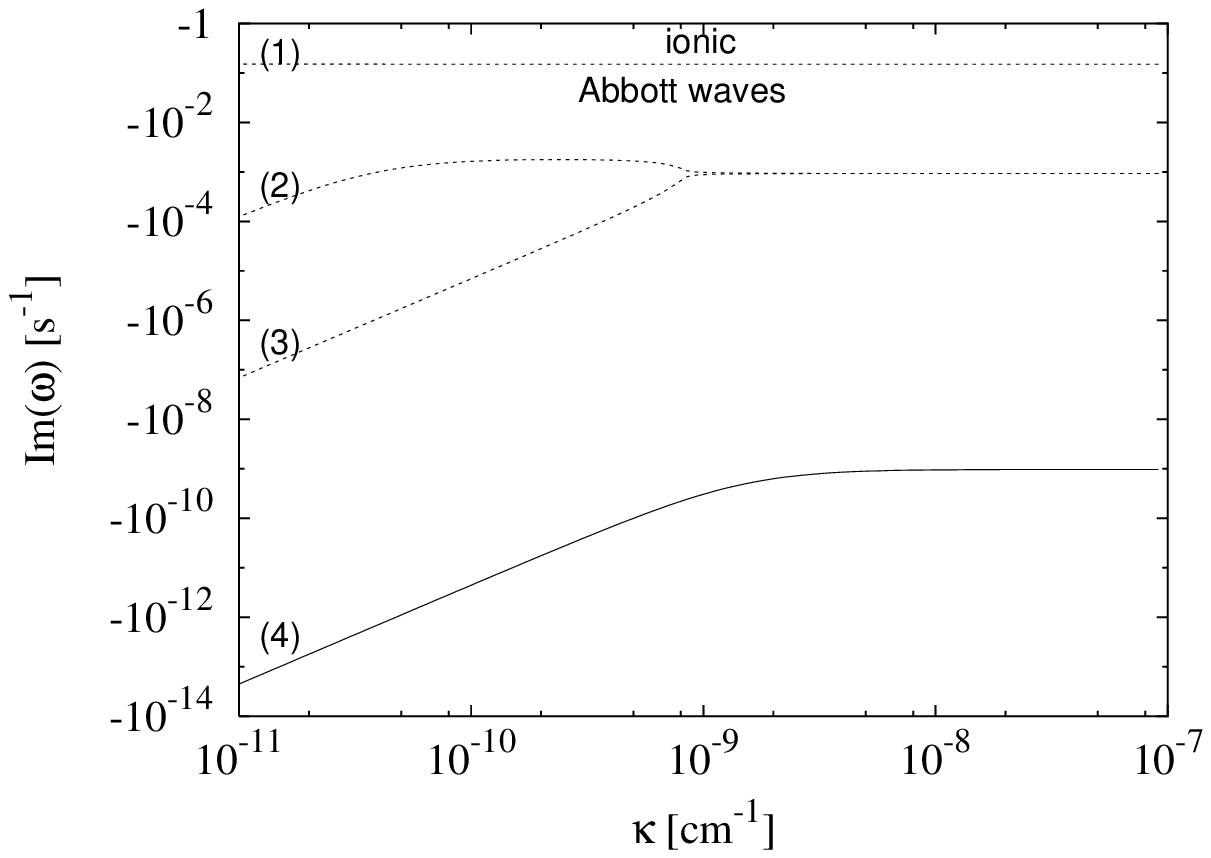}}
\caption[]{Dispersion relation for the point $r=2.4\,R_{*}$
in the wind of a B5 star with $\io q=4.8\pr q$.
{\em Upper panel:} Real roots of $\omega$.
Solid line represents a forward wave, dashed ones backward waves.
The relation approximately corresponding to the dispersion relation
$\omega=-\Delta v_0\kappa$  
(see Eq.~(\ref{dispveldv})) is denoted.
{\em Lower panel:} Imaginary roots of $\omega$, all waves lead to an
instability.
As in the upper panel, solid and dashed lines denote forward and
backward waves, respectively.
Real and imaginary parts of individual roots are denoted using the same
numbers.
}
\label{dispb5}
\end{figure}

Second we study a model of a B5 star with artificially enhanced effect
of friction ($\io q=4.8\pr q$, see KK0) where the multicomponent nature
of the flow plays an important role and leads to lower outflow velocity
than in a one-component case.
Here we study the dispersion relation at the point $r=2.4\,R_{*}$ where
the drift velocity between both components exceeds the value
corresponding to the maximum of Chandrasekhar function and, therefore,
$G'(\Delta v_0)<0$ (see Fig.~\ref{b5chandra}).

The dispersion relations displayed in the Fig.~\ref{dispb5}
substantially differs from the previous case.
Consistently with simplified considerations in the
Sect.\ref{kapiontvln}, the imaginary parts of all roots are negative
and therefore the flow is unstable.
The branch with the largest absolute value of both real and imaginary
parts corresponds to the ionic Abbott waves.
According to the Eq.~(\ref{enabb}), the real part of this branch depends
on $\kappa$ linearly whereas the imaginary part does not depend on
$\kappa$.

A natural question arises: Where have the two-component Abbott waves
disappeared?
However, the simplified calculations yielding the two-component Abbott
waves (see Sect.\ref{kapabb}) are not valid in this case, because the
velocity difference $\Delta v_0$ cannot be simply neglected.
On the other hand, the dispersion relation (\ref{disobec}) can be
approximately fulfilled if the term $\omega+\Delta v_0\kappa$ vanishes.
This conclusion can be simply justified because the dispersion relation
(\ref{disobec}) can be rewritten as
$$A_{\mathrm{pp}}A_{\mathrm{ii}}-A_{\mathrm{pi}}A_{\mathrm{ip}}=0$$ 
and when $\omega+\Delta v_0\kappa=0$ then the $A_{\mathrm{pi}}$ term
vanishes identically and $A_{\mathrm{pp}}$ vanishes if the pressure term
$\pr{a}^2\kappa^2$ is negligible (see Eqs.~(\ref{app},\ref{api})).
The linear dispersion relation
\begin{equation}
\label{dispveldv}
\omega=-\Delta v_0\kappa
\end{equation}
is visible in the Fig.~\ref{dispb5} (note that
$\Delta v_0\approx10^6\,\mathrm{cm}\,\mathrm{s}^{-1}$).

\subsection{On the stability of the B5 star wind model}

\begin{figure}[b]
\resizebox{\hsize}{!}{\includegraphics{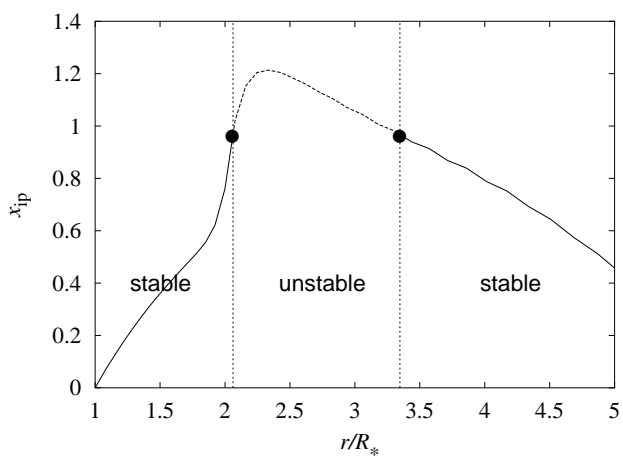}}
\caption[]{
The stability of the wind of a B5 star with $\io q=4.8\pr q$.
At the base of the wind 
$x_{\mathrm{ip}}\lesssim0.97$
(note, that the value
$x_{\mathrm{ip}}\approx 0.97$ 
approximately corresponds to the maximum of Chandrasekhar
function, see Fig.~\ref{b5chandra}) and the wind is stable.
In the outer parts the density decreases, Chandrasekhar function reaches
its maximum and 
$x_{\mathrm{ip}}\gtrsim0.97$,
the wind is not stable.
In the outermost parts of the wind is 
$x_{\mathrm{ip}}\lesssim0.97$
again and
the wind is stable.}
\label{b5stab}
\end{figure}

Nevertheless, the analysis presented in the Section \ref{kapnesta} does
not mean that the {\em whole} wind is unstable for the above mentioned
two-component instability.
Since stability depends on the sign of $G'(\Delta v_0)$ which varies
throughout the flow, there are regions where the wind is stable and
regions where the wind is unstable.
This situation is depicted in the Figure~\ref{b5stab}.

Near the stellar surface the wind is well coupled
($\omega/k\gg \Delta v_0$),
the Chandrasekhar function is below its maximum (as a function of the
velocity difference), $G'(\Delta v_0)>0$ and the wind is stable (see
Eq.~(\ref{enabb})).
This conclusion was justified by numerical calculations at the point
where $\pr\vr\approx4\pr a$ 
\zm{(obtained dispersion relation resembles that in Fig.~\ref{dispepsori}).}

Downstream is the wind accelerated, its density decreases and the
velocity difference $\Delta v_0$ between both components increases.
However, such low density wind reaches the point where the maximum of
the Chandrasekhar function is reached and the absorbing component is not
able to accelerate the passive component sufficiently.
Due to the functional dependence of the radiative force the velocity
gradient of both components decreases (see KK0 and the discussion in
OP), and, in addition, $G'(\Delta v_0)<0$.
As has already been shown in the Section \ref{kapnesta} (see
Fig.~\ref{dispb5}), the flow is unstable in this region.

In the outermost parts of the wind the relation $G'(\Delta v_0)>0$ holds
again and clearly, the wind is stable there.
However, because the wind is unstable upstream where $G'(\Delta v_0)<0$,
the instabilities from that unstable region may disseminate and
influence the stability of the outermost parts of the wind.
Such effects would be studied using hydrodynamical simulations.

Summarizing, there are three regions in the wind of this B5 star, namely
the innermost and outermost parts of the wind are stable against
perturbations larger than the Sobolev length, whilst the ``middle''
part of the wind is unstable (see Fig.~\ref{b5stab}).

The new slow waves that appeared in the stable solution for a dense wind
of an O star (cf. Section \ref{Owaves}) are also present in the stable
parts of the wind of a B5 star.
For the inner region of stability these waves may be both forward and
backward with a special case of 
%Kr12: standing 
\zm{"frozen-in"}
waves for
$\kappa=\kappa_0$ where $\Re(\omega)=0$.
Similarly to the case of an O star,
the value of $\kappa_0$ 
\zm{decreases}
for increasing radii.
These waves are also present in the outer stability region, but in this
case these waves are purely forward.

\subsection{The stability of a B5 star model with
normal friction}

Finally, we studied the stability of a wind model of B5 star with 
$\io q= 2\pr q$, which is a bit closer to real winds of B5 stars,
at the point where $\io\vr\approx 290\,\mathrm{km}\,\mathrm{s}^{-1}$.
At this specific point is $x_{\mathrm{ip}}\approx1.4$ and thus, the flow
is unstable there.
The results of this calculation
\zm{are similar to
those}
displayed in the 
\zm{Fig.~\ref{dispb5}.}
All imaginary roots are negative, the flow is unstable at this
selected point.
The analysis presented in the preceding section is valid also for this
case.

\section{Conclusions}

We showed that two-component isothermal radiatively driven stellar
wind is unstable in the case when friction affects the overall structure
of the wind, i.e. when the drift velocity between both components is
sufficiently large.
Strictly speaking, the wind is unstable if the argument of Chandrasekhar
function is larger than the value for which the maximum of Chandrasekhar
function is reached.
\zm{For this case the so-called ion runaway instability occurs}.
In the opposite case, when the wind is well coupled,
the Abbott waves (for large scale perturbations) in the wind are stable.
Putting these two cases together, stationary wind solutions obtained by
KK0 for the case when the drift velocity increases to such values that
the Chandrasekhar function passes through the point with its maximum
value (i.e. it is decreasing and its derivative is negative) {\em are
not stable}.
Note that the region of instability falls within the region of the
abrupt decrease of the velocity gradient in the solution found by KK0.

In the real case the obtained large growth rate of the instability will
be probably reduced by the energy dissipation via frictional heating.
Nevertheless, the inclusion of frictional heating will probably not
alter the presence and the overall picture of instability for high drift
speeds.
\zm{However, it is not clear whether the ionic component escapes the
star separately (as proposed Springmann \& Pauldrach \cite{treni}),
because the ionic Abbott waves have only modest spatial growth rate (see
OP).
Hydrodynamical simulations are necessary to resolve this problem.
However, it is questionable whether the Sobolev approximation may be
used for such calculations.
The Sobolev approximation was at the edge of its validity during the
calculations of KK0.
If the ionic component leaves the star separately, then the Sobolev
approximation can be used thanks to a large velocity gradient of
absorbing ions after the decoupling (see KKI).
On the other hand, if the decoupling process is more complex with modest
velocity gradients,
we may come beyond the region of validity of the Sobolev approximation.}

More complete dispersion relations for two-component flow should
be derived with the inclusion of the short-wavelength perturbations.
In the one-component case these perturbations lead to the well-known
radiatively driven wind instability (MacGregor et al. \cite{nesmac},
Carlberg \cite{nesta}, Owocki \& Rybicki \cite{ornest}).
When such perturbations on a scale below the Sobolev length are
included, then the onset of a two-component instability occurs
even below the maximum of Chandrasekhar function (see OP).

The existence of the two-component instability could also lead to
frictional heating of the wind up to the temperatures of the order of
$10^6\,\mathrm{K}$ (cf. KKII).
Such heating could explain enhanced X-ray activity of many B-stars.
Similar mechanism of X-ray generation was also proposed by Porter \&
Drew (\cite{iontdisk}) and OP.
Unfortunately, in the latest numerical models of a three-component
nonisothermal wind of B stars the velocity difference is too low
to allow for such ion runaway instability (except for extremely low
density winds).
On the other hand these results are based on a bit artificial dependence
of the radiative force on the temperature
and a more advanced calculations can alter this result.

In a case of a wind where two component effects are not important (an
O star wind), the two-component stability analysis enabled us to find
more types of waves than the classical one-component analysis of Abbott
(\cite{abb}).
In addition to the original stable Abbott waves, there exist heavily
damped ionic waves and very slow waves with very weak damping.
The latter waves may move in both directions with respect of the wind or
they may be
\zm{static}
in the comoving fluid frame.
Such slow waves resemble
\zm{"frozen-in" wavy pattern}
of the dimension of approximately
$\sim 10^7 \mathrm{cm}$.

\begin{acknowledgements}
The authors would like to thank to Dr. Achim Feldmeier for his
valuable comments on the manuscript.
This research has made use of NASA's Astrophysics Data System Abstract
Service (\cite{ADS1}, \cite{ADS2}, \cite{ADS3}, \cite{ADS4}). 
This work was supported by a grant GA \v{C}R 205/01/0656,
by a grant GA AV \v{C}R A3003805,
and by projects K2043105 and Z1003909.
\end{acknowledgements}


\begin{thebibliography}{}

\bibitem[1980]{abb} Abbott, D. C., 1980, ApJ 242, 1183
\bibitem[1982]{abpar} Abbott, D. C., 1982, ApJ 259, 282
\bibitem[Accomazzi et al. 2000]{ADS3} Accomazzi, A., Eichhorn, G.,
              Kurtz, M. J., Grant, C. S., Murray, S. S., 2000, A\&AS
              143, 85
\bibitem[1995]{babela} Babel, J., 1995, A\&A 301, 823
\bibitem[1980]{nesta} Carlberg, R. G., 1980, ApJ 241, 1131
\bibitem[1975]{cak} Castor, J. I., Abbott, D. C., Klein, R. I., 1975,
              ApJ 195, 157 (CAK)
\bibitem[1959]{dreicer1} Dreicer, H., 1959, Phys. Rev. 115, 238
\bibitem[Eichhorn et al. 2000]{ADS2} Eichhorn, G., Kurtz, M. J.,
              Accomazzi, A., Grant, C. S., Murray, S. S., 2000, A\&AS
              143, 61
%Ku14: \bibitem[1998]{feld98} Feldmeier, A., A\&A 332, 245
%      chybel rok, over to, prosim
\bibitem[1998]{feld98} Feldmeier, A., 1998, A\&A 332, 245
%
\bibitem[2000]{feslo00} Feldmeier, A.,  Shlosman, I., 2000, ApJ 532,
              L125
\bibitem[2001]{feslo01} Feldmeier, A.,  Shlosman, I., 2001, ApJ 564, 385
\bibitem[1986]{fa} Friend, D. B., Abbott, D. C., 1986, ApJ 311, 701
\bibitem[Grant et al. 2000]{ADS4} Grant, C. S., Accomazzi, A., Eichhorn,
              G., Kurtz, M. J., Murray, S. S., 2000, A\&AS 143, 111
\bibitem[1970]{jetra} Jenkins, M. A., Traub, J. F., 1970, SIAM J. Numer.
              Anal. 7, 545
\bibitem[2001]{disertacka}  Krti\v{c}ka, J., 2001, Ph.D. thesis, Masaryk
              University Brno
\bibitem[2000]{kk} Krti\v{c}ka, J., Kub\' at, J., 2000, A\&A 359, 983
              (KK0)
\bibitem[2001a]{kki} Krti\v{c}ka, J., Kub\' at, J., 2001a, A\&A 369, 222
              (KKI)
\bibitem[2001b]{kkii} Krti\v{c}ka, J., Kub\' at, J., 2001b, A\&A 377,
              175 (KKII)
\bibitem[Kurtz et al. 2000]{ADS1} Kurtz, M. J., Eichhorn, G., Accomazzi,
              A., Grant, C. S., Murray, S. S., Watson, J. M., 2000,
              A\&AS 143, 41
\bibitem[1982]{L82} Lucy, L. B., 1982, ApJ 255, 286
\bibitem[1984]{L84} Lucy, L. B., 1984, ApJ 284, 351
\bibitem[1970]{LS70} Lucy, L. B., Solomon, P. M., 1970, ApJ 159, 879
\bibitem[1980]{luw} Lucy, L. B., White, R. L., 1980, ApJ 241, 300
\bibitem[1979]{nesmac} MacGregor, K. B., Hartmann, L., Raymond, J. C.,
              1979, ApJ 231, 514
\bibitem[1992]{aetso} Owocki, S. P., 1992, in The Atmospheres of
              Early-Type Stars, U.Heber \& C.S.Jeffery eds., Lecture
              Notes in Physics Vol. 401, Springer Verlag Berlin, p. 393
\bibitem[1999]{owco} Owocki, S. P.,  Cohen, D. H., 1999, ApJ 520, 833
%Kr14: \bibitem[2001]{op} Owocki, S. P., Puls, J., 2001, ApJ, in press (OP)
\bibitem[2002]{op} Owocki, S. P., Puls, J., 2002, ApJ, in press (OP)
%
\bibitem[1984]{ornest} Owocki, S. P., Rybicki, G. B., 1984, ApJ 284, 337
\bibitem[1985]{or2} Owocki, S. P., Rybicki, G. B., 1985, ApJ 299, 265
\bibitem[1986]{or3} Owocki, S. P., Rybicki, G. B., 1986, ApJ 309, 127
\bibitem[1991]{or5} Owocki, S. P., Rybicki, G. B., 1991, ApJ 368, 261
\bibitem[1986]{ppk} Pauldrach, A., Puls, J., Kudritzki, R. P., 1986,
              A\&A 164, 86
\bibitem[1995]{iontdisk} Porter, J. M., Drew, J. E., 1995, A\&A 296, 761
\bibitem[1987]{rybinst} Rybicki, G. B., 1987, in Instabilities in
              Luminous Early Type Stars, H. J. G. L. M. Lamers \& C. W.
              H. de Loore eds., D. Reidel Publ. Comp., Dordrecht, p.175
\bibitem[1990]{or4} Rybicki, G. B., Owocki, S. P., Castor, J. I., 1990,
              ApJ 349, 274
\bibitem[1992]{treni} Springmann, U. W. E., Pauldrach, A. W. A., 1992,
                     A\&A 262, 515
\end{thebibliography}
\end{document}